**TITLE**

Gate-Tunable Optical Nonlinearities and Extinction in Graphene/LaAlO$_3$/SrTiO$_3$ Nanostructures

**AUTHORS**

Erin Sheridan[1,2], Lu Chen[1,2], Jianan Li[1,2], Qing Guo[1,2], Ki-Tae Eom[3], Hyungwoo Lee[3], Jung-Woo Lee[3], Chang-Beom Eom[3], Patrick Irvin[1,2], Jeremy Levy[1,2] *

[1] *Department of Physics and Astronomy, University of Pittsburgh, Pittsburgh, PA 15260, USA*

[2] *Pittsburgh Quantum Institute, Pittsburgh, PA, 15260, USA*

[3] *Department of Materials Science and Engineering, University of Wisconsin-Madison, Madison, WI 53076, USA*

*\* Corresponding author (jlevy@pitt.edu)*

**ABSTRACT**

Pristine, undoped graphene has a constant absorption of 2.3 % across the visible to near-infrared (VIS-NIR) region of the electromagnetic spectrum. Under certain conditions, such as nanostructuring and intense gating, graphene can interact more robustly with VIS-NIR light and exhibit a large nonlinear optical response. Here, we explore the optical properties of graphene/LaAlO$_3$/SrTiO$_3$ nanostructures, where nanojunctions formed at the LaAlO$_3$/SrTiO$_3$ interface enable large (~$10^8$ V/m) electric fields to be applied to graphene over a scale of ~10 nm. Upon illumination with ultrafast VIS-NIR light, graphene/LaAlO$_3$/SrTiO$_3$ nanostructures produce broadband THz emission as well as a sum-frequency generated (SFG) response. Strong spectrally sharp, gate-tunable extinction features (>99.99%) are observed in both the VIS-NIR and SFG regions alongside significant intensification of the nonlinear response. The observed gate-tunable strong graphene-light interaction and nonlinear optical response are of fundamental interest and open the way for future exploitation in graphene-based optical devices.

**INTRODUCTION**

Graphene[1], a monatomic layer of *sp$^2$* carbon atoms, exhibits many unique optical properties[2] that make the material desirable for a variety of quantum and optical device applications[3]. The absorption for pristine, undoped graphene is defined by the fine structure constant $\alpha$, and is constant at $\pi\alpha \approx 2.3\%$ across visible to near-infrared (VIS-NIR) frequencies[4]. The chemical potential $\mu$, which can be tuned via electrostatic gating or chemical doping with respect to the charge-neutrality point (CNP), has only a modest effect on its optical absorption[5,6]. While this flat and weak response may be beneficial for some applications (e.g, transparent ultrathin conductors[7]), the low absorption is limiting for other applications, such as photodetectors[8].

For light in the mid-infrared to far-infrared region, graphene hosts a plasmonic response[9], and as a result, strong graphene absorption can be induced in this frequency range by gating the plasmons[10-12].

Because graphene typically lacks a plasmonic response in the VIS-NIR regime, such behavior is difficult to achieve at higher frequencies[13]. However, the interaction between graphene and VIS-NIR light can be enhanced by creating graphene-based metamaterials or surfaces in which the CNP is modulated at the nanoscale, for example, using AFM[14] or STM[15], by creating arrays of graphene nanodisks or nanoribbons[10-12,16], or by placing graphene near plasmonic metasurfaces or nanoscale metal gratings[17-20].

Recently, a technique to control the CNP of graphene—both reversibly and locally—has been developed using graphene integrated with $LaAlO_3$/$SrTiO_3$ (LAO/STO) heterostructures[21,22]. LAO/STO has a tunable conductive interface[23] with a variety of interesting physical properties[24]. When the LAO thickness is close to the critical thickness for a metal-insulator transition, ~3-4 unit cells[25], the conductivity of the LAO/STO interface can be controlled using conductive atomic force microscope (c-AFM) lithography[14,26]. A wide range of optoelectronic devices can be fabricated at the LAO/STO interface in this fashion, such as a 10 nm-scale photodetector[27] and nanoscale, terahertz (THz) sources and detectors[28,29] with a bandwidth of more than 100 THz. LAO/STO nanostructures can be placed within two nanometers of an active graphene device and used, for example, to create reconfigurable edge channels in graphene[22].

**EXPERIMENTAL SETUP**

The nonlinear optical properties of graphene/LAO/STO (G/LAO/STO) nanojunctions (illustrated in **Figure 1(b)**) are measured through a broadband THz spectroscopy technique (**Figure 2**) that takes advantage of strong optical nonlinearities in STO[28,29,30]. The G/LAO/STO nanojunctions are created using c-AFM lithography, described in detail elsewhere[14] and summarized in the **Materials and Methods** section. A nanojunction (**Figure 1(b)**) consists of a conducting LAO/STO nanowire with a nanoscale (~10 nm) insulating gap. The nanojunction is defined directly underneath a single layer of graphene[21] that is patterned into the shape of a Hall bar (**Figure 1(a)**). Because the nanojunction is only 1.2 nm below the graphene, there is a strong electrostatic interaction between the graphene and the nanojunction. Details of sample growth and graphene patterning and integration can be found in the **Materials and Methods** section.

Here we measure the effects of local electrostatic gates ($V_{NW}$, $V_{SD}$) and a global gate ($V_{gr}$) (defined below) on the optical response of the G/LAO/STO nanostructure. The nanostructure is illuminated by ultrafast pulses and its response is measured as a function of the time delay $\tau$ between two pulses[28,29]. A detailed schematic drawing of the optical setup is shown in **Figure 2**. Ultrafast pulses emerge from a sub-7-fs Ti:Sapphire oscillator with a repetition rate of 80 MHz. Pulses are sent through an optical pulse shaper and into a compact Michelson interferometer before being focused to a diffraction-limited spot centered on the nanojunction. The time-averaged input excitation power from the Ti:Sapphire laser varies from 5-15 µW. During the measurement, a delay line for one of the beams is scanned from negative to positive time delay $\tau$ while the induced photovoltage across the LAO/STO nanojunction $\Delta V(\tau)$ (e.g., **Figure 1(c,d)**) is recorded. Delay scans are repeated 30-50 times to increase the signal/noise ratio. $\Delta V(\tau)$ is Fourier transformed with respect to $\tau$ to yield a power spectrum magnitude (log scale) versus frequency $\Omega$ (**Figure 1(e)**).

For photovoltage measurements, two electrodes, labeled **S** and **D** in **Figure 1(a),** are used to apply a DC bias voltage $V_{SD}$ across the LAO/STO nanojunction. Voltage-sensing leads (**V+** and **V−**) are used to measure the photovoltage change across the LAO/STO nanojunction that is induced by ultrafast laser pulses. The biased nanojunction produces a local region with a large DC bias field, of order $1 \times 10^8$ V/m for a typical bias $V_{SD} = 1$ V.

LAO/STO nanojunctions without graphene have been shown to locally generate and detect THz emission with >100 THz bandwidth via the third order nonlinear optical process in STO[31] (see **Supplementary Figures S9-11** for an example control experiment). $V_{SD}$ creates a quasi-static electric field $\vec{E}_{bias}$ across the junction that is highly confined in space to ~10 nm, while input optical fields $\vec{E}_{opt}(\omega_1), \vec{E}_{opt}(\omega_2)$ are sharply peaked in the time domain. The three electric fields mix to generate the nonlinear response of the nanojunction at the difference $(\omega_1 - \omega_2)$ and sum $(\omega_1 + \omega_2)$ frequencies. Alternatively, the third-order nonlinear susceptibility is converted, via $V_{SD}$, into a local second order nonlinear susceptibility at the site of the nanojunction[29]. The difference-frequency generated response ("DFG") ranges from 0-100 THz, and the sum-frequency generated response ("SFG") extends from about 700-850 THz. The response at the fundamental excitation, which exists primarily in the near-infrared but extends into the visible range 340-450 THz (1.4-1.9 eV), constitutes the linear response ("LNR") of the junction to the input laser excitation.

Time-domain photovoltage measurements of G/LAO/STO nanojunctions are taken under a variety of experimental conditions at temperatures ranging from 5-50 K. There are three electrical gates that are tuned: $V_{gr}, V_{NW}, V_{SD}$. $V_{gr}$ is a "global" bias with respect to the STO back gate, while $V_{NW}$ represents the local nanowire bias (common mode), and $V_{SD}$ is the differential bias across the LAO/STO junction (differential mode). For all experiments, the back gate to the sample $V_{bg}$ is grounded. Other degrees of freedom explored include the average optical power and linear polarization (see **Supplementary Figure S7**). Here we focus mainly on the gate-dependent measurements, but show some examples of the dependence of the G/LAO/STO nanostructure response on optical parameters. The photovoltage $\Delta V(\tau)$ is measured at different gate values, while the four-terminal resistance of the graphene $R_{G,4T}$ is measured at the same conditions. The **Supplementary Information** provides a description of the four-terminal resistance measurement and accompanying diagram (**Supplementary Figure S2**.)

**RESULTS**

The graphene chemical potential is first tuned by applying a DC offset to the global graphene gate $V_{gr}$ with respect to ground to graphene electrodes labeled **V_gr** in **Figure 1(a)**. The results of a $V_{gr}$-dependent experiment are summarized in **Figure 1(c-e)** and described in more detail below. **Figure 1(c)** shows the time domain measurement at $V_{gr} = 0$, which is similar to the response from LAO/STO nanojunctions without graphene. When $V_{gr} = -0.3$ V, shown in **Figure 1(d)**, the time-domain signal changes significantly. In the corresponding power spectrum (**Figure 1(e)**), a sharp, four-order-of-magnitude extinction feature is revealed within the LNR spectral region at 380 THz (1.57 eV). Additionally, the nonlinear response of the G/LAO/STO nanostructure is enhanced at $V_{gr} = -0.3$ V as compared to the power spectrum of the $V_{gr} = 0$ V response. Experiments performed on control LAO/STO nanojunctions in the absence of graphene do not show these extinction features (see **Supplementary Figure S9-11** for an example control experiment).

DEPENDENCE OF TIME-DOMAIN PHOTOVOLTAGE RESPONSE ON GLOBAL GRAPHENE GATE

As summarized above, when the graphene chemical potential is tuned via $V_{gr}$, the response of the G/LAO/STO nanojunction changes dramatically. Two-terminal and four-terminal graphene conductance plots vs. $V_{gr}$ for this experiment are available in **Supplementary Figure S3**. This method of changing the chemical potential does not clearly reveal the graphene CNP, as compared to the $V_{NW}$ gating scheme described below. Four representative time-domain photovoltage traces and their corresponding power spectra are shown in **Figure 3** for different $V_{gr}$ values at $T = 10$ K and $V_{SD} = -1$ V. The extinction feature only occurs over a narrow gate range, $V_{gr} = [-0.5 \text{ V}, -0.2 \text{ V}]$, and can be observed in the time domain signals in **Figure 3(a)**, **(c)**. Additionally, the nonlinear (DFG, SFG) regions of the power spectra for these time domain signals is quite large. In comparison, the nonlinear regions of the power spectra in **Figure 3(f)**, **(h)**, where no extinction is observed, have lower amplitudes.

To further examine the enhancement of the nonlinear response, integrals of the DFG, LNR and SFG regions of the power spectra are taken as a function of $V_{gr}$ and plotted in **Figure 4(a-c)**. The nonlinear response of the G/LAO/STO nanostructure is maximal at the $V_{gr}$ value where the sharp extinction line in the LNR range appears, whereas the LNR response is maximal at $V_{gr} = 0$ V. A comparison of the DFG and SFG integrals (**Figure 4(d)**) shows a power law dependence $SFG = a(DFG^m)$ with the exponent $m = 0.79 \pm 0.056$. Ideally $m \approx 1$, which would show that the integrals of the SFG and DFG nonlinear signals have a 1:1 correspondence. However, the DFG and SFG signals still do appear to have a linear relationship with $m \approx 0.8$. The complete data set for this experiment is available in **Supplementary Figures S4-5**.

NANOWIRE GATE DEPENDENCE

In a separate experiment, the graphene is gated by applying a DC offset to both the source **S** and drain **D** of the LAO/STO nanojunction (i.e. common mode). Since the conducting nanojunction structure is separated from the graphene by a ~1.2 nm thick dielectric LAO layer, a DC offset applied to the nanojunction ($V_{NW}$) acts as a gate directly beneath the graphene. $V_{NW}$ allows for a more local gating effect on the graphene chemical potential.

Five representative time domain signals and power spectra for different $V_{NW}$ values are shown in **Figure 5**, with $T = 45$ K and $V_{SD} = -2$ V. A sharp extinction line appears in the LNR region at 375 THz when $V_{NW} = -1.5$ V. Remarkably, an extinction line also appears in the SFG response at 760 THz when $V_{NW} = 1$ V. The graphene four-terminal resistance $R_{G,4T}$ is plotted as a function of $V_{NW}$ in **Figure 6(a)**, where the CNP is clearly visible at $V_{NW} = 1.7$ V. It is notable that, in addition to the CNP at $V_{NW} = 1.7$ V, there is a second local maximum in $R_{G,4T}$ at $V_{NW} = -2$ V. The second local maximum may be a signature of inhomogeneous doping in the graphene by the nanojunction, as described further in the **Discussion** section.

Integrals of the relevant regions of the power spectra versus $V_{NW}$ are plotted in **Figure 6(b-d)**. As with the $V_{gr}$ dependence, a local maximum in the DFG and SFG integrals occurs when the extinction feature appears in the LNR response, at about $V_{NW} = -1.5$ V. Additionally, both nonlinear signals have local maxima near the CNP of graphene ($V_{NW} = 1.7$ V). In contrast, the integral of the LNR signal appears to have a maximum only near the CNP.

LAO/STO JUNCTION DIFFERENTIAL MODE DEPENDENCE

To further explore the electrostatic interaction between the LAO/STO nanojunction and the graphene, the graphene global gate $V_{gr}$ is held constant at $-25$ mV, with $V_{NW} = 0$, and the nanojunction differential mode $V_{SD}$ is varied at $T = 10$ K. As shown in the power spectra plotted in **Figure 7(a)**, the $V_{SD}$ dependence of the VIS-NIR extinction feature is moderate. As $V_{SD}$ increases, the observed extinction feature in the LNR response becomes spectrally sharper and stronger before returning to a broader and weaker form. This effect can be observed clearly in the integral of the power spectra from 420-430 THz at each $V_{SD}$ value in **Figure 7(c),** where the decreasing then increasing integral value reflects the evolving amplitude of the extinction feature. Additionally, the frequency of the extinction feature shifts very slightly with $V_{SD}$. Therefore, unlike $V_{gr}$ or $V_{NW}$, $V_{SD}$ can more easily be used as a fine-tuning parameter to precisely control the amplitude and frequency of extinction features.

OPTICAL POWER DEPENDENCE

Power-dependent experiments reveal a tuning ability similar to $V_{SD}$. Time domain photovoltage measurements are obtained at different input powers ranging from $7 - 11$ µW at $T = 10$ K, $V_{SD} = -1.25$ V and $V_{NW} = 0$ (shown in **Figure 7(b)**). As the input power to the G/LAO/STO nanojunction decreases, the VIS-NIR sharp extinction line decreases in amplitude. The decrease in the extinction is reflected as an increase in the integral of the power spectrum from 385-415 THz in **Figure 7(d)** as the optical power decreases. Although the $V_{SD}$ and optical power dependences of the extinction feature appear to be similar and interrelated (see **Supplementary Figure S6**), the complex interplay of various optical and electronic effects on the G/LAO/STO nanostructure prohibit the formulation of a specific scaling law.

**DISCUSSION**

The observed >99.99% extinction features in the VIS-NIR in G/LAO/STO nanostructures are highly unusual given graphene's typical optical behavior. When the energy of the input radiation $\hbar\omega$ is greater than or equal to the Fermi energy $2\mu$ of the graphene, $\hbar\omega \gtrsim 2\mu$, the conductivity of bulk Dirac fermions is independent of frequency, and therefore the optical absorption of graphene is expected to be constant in that region[17]. Spectral selectivity, or increased extinction of light at a particular frequency, in graphene-based structures has been reported in the near-infrared or visible range [11,17,18,20,32,33] and has been ascribed to magnetic polaritons (MP), surface plasmon polaritons (SPP), or resonances of plasmonic nanostructures or underlying metallic or dielectric cavities. These plasmon resonances exist independently of the graphene, which serves only to intensify or tune an absorption resonance that is already there.

One possible interpretation of our results is that the G/LAO/STO nanostructure, with its extremely confined electric field at the nanoscale as depicted in **Figure 1(b)**, may act as a highly doped graphene nanoribbon, nanodisk or quantum dot-like structure. As discussed in Refs. [16,34], extended graphene sheets can be inhomogeneously doped by patterning them with an underlying back gate similar in nature to the LAO/STO nanojunction presented here. The generation of confined VIS-NIR graphene plasmons, or trapped plasmons in a *p-n* junction line, is plausible in these structures and can lead to absorption of light at the plasmon energy[16,34].

In support of this picture, $R_{G,4T}$ in **Figure 7(a)** shows an additional local maximum away from the CNP, which may be a signature of inhomogeneous doping in the extended graphene sheet by the biased LAO/STO nanojunction. The separation between two Dirac point features in $R_{G,4T}$ as a function of $V_{SD}$ is plotted in **Supplementary Figure S12**. In general, as $V_{SD}$ increases, the separation between the two observed Dirac features increases, suggesting the presence of inhomogeneous doping of the graphene by the biased nanojunction. (More details available in **Supplementary Information**.) Furthermore, nanostructured graphene has been predicted to exhibit strong, plasmon-enhanced nonlinear optical behavior, which could explain the enhanced nonlinear optical response of the G/LAO/STO structure when the extinction feature appears[35].

While it is true that a slight shift of the graphene CNP occurs due to LAO/STO patterning, the resultant change in carrier density ($\Delta n = 7 \times 10^{11}$ cm$^{-2}$ )[22] is negligible compared to the effects of the biased nanojunction on the chemical potential. For $V_{SD} = 1$ V, we estimate a change in carrier density of $\Delta n = 2.8 \times 10^{13}$ cm$^{-2}$ and a corresponding chemical potential change of $\Delta \mu = \hbar v_F \sqrt{\pi \Delta n} = 0.62$ eV. Details of this estimate can be found in the **Materials & Methods** section. However, it is important to note that the LAO/STO nanojunction affects the graphene charge carriers in a highly non-uniform fashion. The G/LAO/STO nanostructure is subject to strong electric fields that can create a large local dipole, and the effect of this gating on the overall graphene chemical potential is difficult to calculate.

The strong ~$10^8$ V/m localized electric field across the G/LAO/STO nanostructure may also lead one to consider nonlinear plasmonic processes, such as multi-plasmon absorption, as discussed by Jablan and Chang[36]. Since we observe a strong enhancement of the optical nonlinearity of the G/LAO/STO nanostructure, we also must take into account graphene's large intrinsic optical nonlinearities[37-41]. Finally, one must keep in mind that in these experiments the G/LAO/STO nanostructure is excited with nearly transform-limited ultrafast pulses (see **Materials & Methods** for more details), and such an excitation is known to create non-equilibrium hot electrons and transient plasmons in graphene[42-44].

These and other effects must be considered in a rigorous theoretical model of the G/LAO/STO nanostructure response.

**CONCLUSIONS**

Our demonstrated ability to induce >99.99% narrow band extinction of light across a broad range of VIS-NIR frequencies (see **Supplementary Figure S8** for more examples) in a G/LAO/STO nanostructure is highly unusual, considering the low intrinsic absorption of graphene. These sharp extinction lines occur alongside a notably strong enhancement of optical rectification and second harmonic generation in the nanostructure. The G/LAO/STO nanostructure exhibits behavior reminiscent of doped graphene nanoribbon-, nanoisland-, or nanodisk-like structures[35,39,45,46]. In addition to its fundamental significance, this gate-tunable tunable enhanced light-matter interaction promises to strengthen graphene as a candidate material for nanophotonics and quantum optics applications.

**MATERIALS AND METHODS**

LAO/STO SAMPLE GROWTH

The LAO/STO samples are grown via pulsed laser deposition. A thin film (3.4 or 8 unit cells) of LAO is deposited epitaxially on the (001) $TiO_2$-terminated STO substrate at 550 °C and an oxygen pressure of $10^{-3}$ mbar, with its thickness monitored in situ via high-pressure reflection high-energy electron diffraction (RHEED). Electrical contacts to the interface are fabricated via conventional photolithography, where predefined regions are etched via Ar+ ion milling (25 nm) and filled with Ti/Au (4 nm/25 nm). A second layer of Ti/Au is added on top of the LAO surface for wire bonding.

GRAPHENE GROWTH AND PATTERNING

The graphene used in this work is grown from chemical vapor deposition (CVD) on oxygen-free electronic grade copper flattened with a diamond turning machine. The graphene is then coated with the perfluoropolymer Hyflon AD60 and transferred onto the LAO/STO surface using a wet-transfer technique. Graphene is patterned into Hall bars via standard photolithography. The Hyflon coating is removed from graphene with FC-40 after patterning. Particles and contaminants on graphene from wet transfer and photolithography are brushed away using a contact-mode AFM scan sequence. After cleaning, the 4 Å atomic steps of the LAO surface underneath graphene are clearly resolvable, showing that the LAO/graphene interface is clean. The quality of the graphene is comparable to other samples prepared using similar methods, with the mobility $\mu > 10\,000$ $cm^2\,V^{-1}\,s^{-1}$ at 2 K.

C-AFM LITHOGRAPHY

In ambient conditions, a positively biased c-AFM tip dissociates protons from water molecules. During c-AFM lithography, a positively biased AFM tip is scanned along a line in contact mode over the LAO surface to locally charge the LAO surface with protons. These protons locally gate the LAO/STO interface to form conducting nanostructures. Nanowires created using this method have a typical width of 10 nm. A negatively biased AFM tip that is scanned over the conducting regions will remove the adsorbed protons, thereby locally restoring the interface to an insulating state. Using this method, custom, reversible and reconfigurable nanoelectronic and nanophotonic devices can be fabricated.

To create a G/LAO/STO nanostructure, a nanowire is created by scanning the c-AFM tip on top of graphene with a positive voltage, and the nanojunction is created in a region where graphene covers the LAO surface by applying a negative voltage pulse at a point along the nanowire. See **Figure 1 (a, b)** and **Supplementary Figure S1 (b)** for illustrations of the device geometry. Once c-AFM writing is finished, the sample is placed in vacuum in an optical cryostat and cooled to $T$ = 5-50 K.

PULSE SHAPING AND TIME DOMAIN MEASUREMENTS

Ultrafast pulses leave from a Spectra-Physics Rainbow 2 UHP sub-7-fs Ti:Sapphire oscillator with a center wavelength of 800 $\pm$ 20 nm. The ultrafast pulse shaper is a homebuilt system in 4$f$ configuration that is based on a dual-mask spatial light modulator (SLM, Jenoptik SLM-S640d)[47], where wavelengths are spatially separated by a grating and focused onto various pixels of the SLM. Both the amplitude and the phase of the ultrafast pulse can be controlled independently using the dual mask SLM and software that we have built. To change the input power to the sample, the pulse shaper is used to change the amplitude of the input pulses to the G/LAO/STO device. For pulse compensation, the Multiphoton Intrapulse Interference Phase Scan (MIIPS) optimization algorithm is used[48].

Our home-built Michelson interferometer has two arms of approximately equal length. A p-polarized 50/50 ultrafast beam splitter (BS) splits the input pulses into two beams. The reflected beam is normally incident to a plane mirror (PM) that is mounted on a piezoelectric stage (PS), which serves as an optical delay line. The transmitted beam reflects off a plane mirror that is mounted on a mechanical stage, which enables coarse adjustment of the time delay $\tau$. Both beams are recombined by the same beam splitter after normal reflection and subsequently focused onto the nanojunction using a 100x, 0.73 NA objective (OB). During the measurement, the delay line is scanned continuously from negative to positive time delay values. A DC bias voltage $V_{SD}$ is applied to electrode **S** through a 50-Ω-impedance analog output port, while electrode **D** is grounded. The photovoltage, which is the voltage difference, namely, **ΔV = V+−V−**, between the two voltage sensing electrodes, is measured and amplified by a differential voltage amplifier (DVA) with 0.1 MΩ input impedance and recorded as a function of τ.

GRAPHENE CHEMICAL POTENTIAL SHIFT ESTIMATE

To estimate the change in chemical potential for an applied DC bias of $V_{SD} = 1$ V, we employ a capacitive model[11], with assumed capacitance density between the LAO layer and graphene[49] $C = 4.5 \frac{\mu F}{cm^2}$. Using this simple model, a change in carrier density of $\Delta n = \frac{CV_{SD}}{e} = 2.8 \times 10^{13}$ cm$^{-2}$ is calculated. Next, the chemical potential change is estimated using the relation $\Delta \mu(n) = \hbar v_F \sqrt{\pi \Delta n} = 0.62$ eV.

**FUNDING ACKNOWLEDGEMENTS**

This work was supported by ONR (N00014-16-1-3152). E.S. acknowledges support from a National Science Foundation Graduate Research Fellowship Program under Grant No. 1747452. The work at University of Wisconsin-Madison was supported by Air Force Office of Scientific Research FA9550-15-1-0334.

**AUTHOR CONTRIBUTIONS**

E.S. and L.C. performed experiments and processed the data. L.C. designed and built the Michelson interferometer and pulse shaper. E.S. developed the software for the pulse shaper. H.L., J.W.L. and K.T.E. grew the LAO/STO samples. J.Li and Q.G. grew and patterned the graphene on top of LAO/STO. C.B.E., P.I. and J.Levy directed and supervised the project. All co-authors contributed to writing of the manuscript.

**COMPETING FINANCIAL INTERESTS**

The authors declare no competing financial interests.

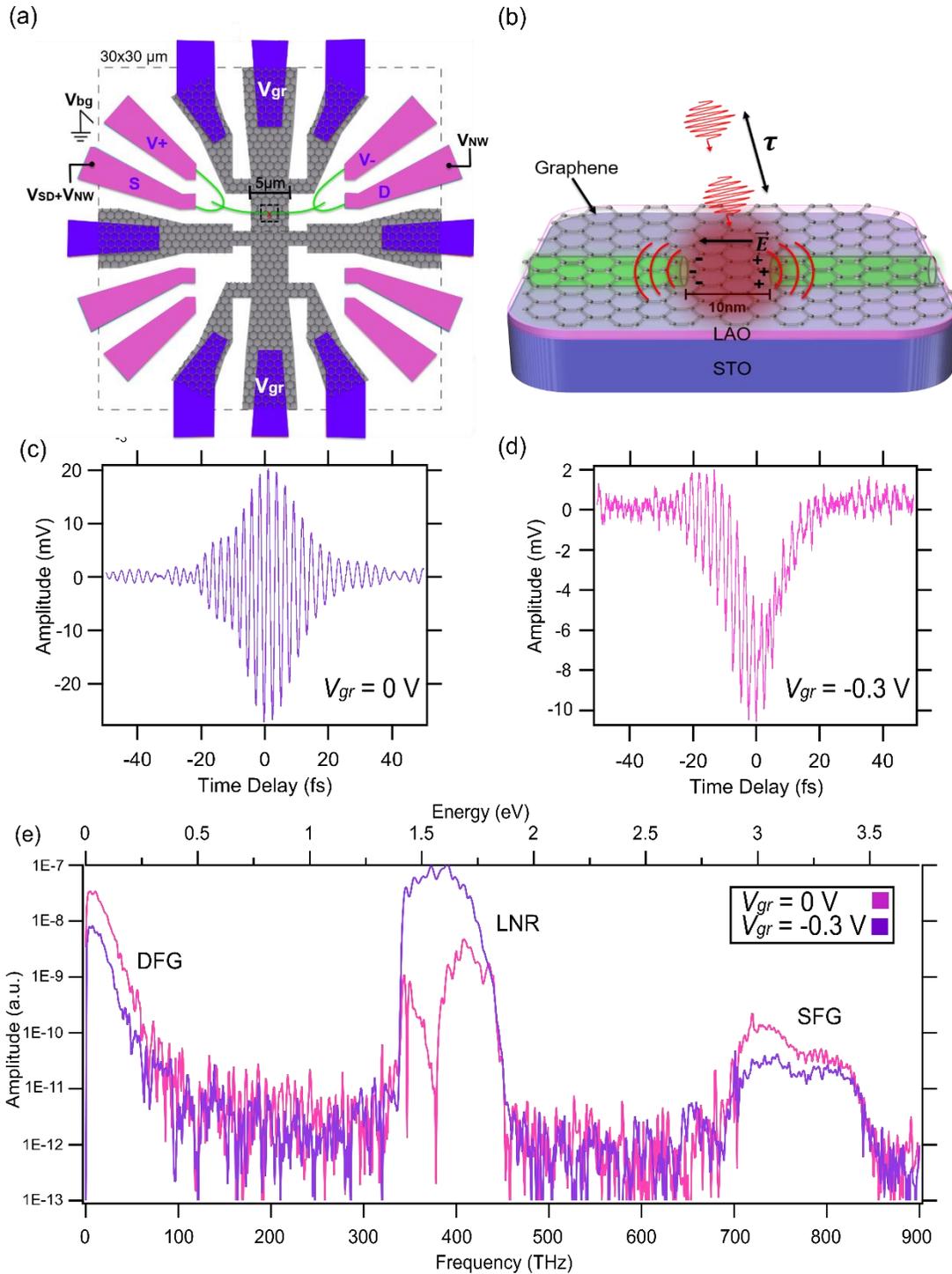

**Figure 1. Summary of experiment and results. (a)** Top-view of G/LAO/STO heterostructure. Purple electrodes denote connections to the graphene Hall bar. Pink electrodes contact the LAO/STO interface for c-AFM lithography and time domain photovoltage measurements. A nanojunction device is sketched across the top half of the 5μm-wide Hall bar. The small black dashed box marks the location of the nanojunction. **(b)** Zoomed-in side view of the biased LAO/STO nanojunction underneath the graphene. Image is not to scale. **(c,d)** Example time domain measurements at $V_{gr} = 0$ V (c) and $V_{gr} = -0.3$ V (d), which shows a striking splitting feature. **(e)** Power spectra of the time domain signals in (c,d). When the time domain signal shows splitting, a sharp extinction line appears in the power spectrum, here at 380 THz. Note an increased DFG and SFG response when sharp extinction is observed.

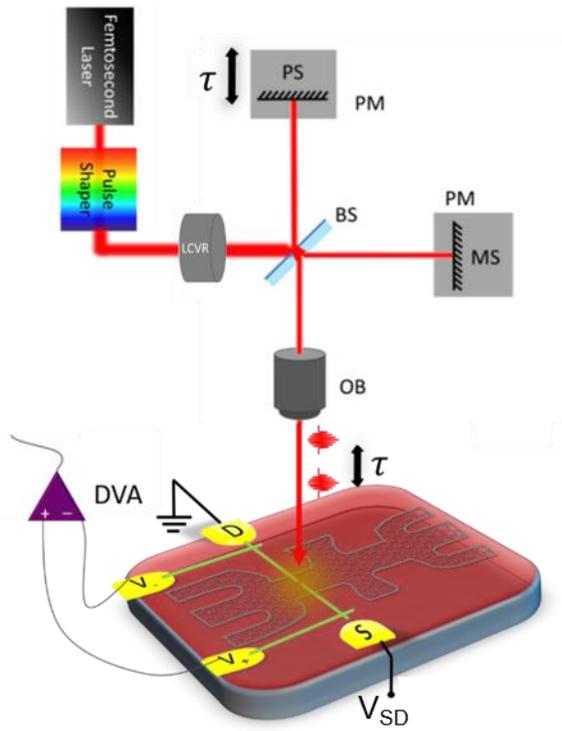

**Figure 2. Optical setup.** Schematic diagram of the optical setup. LCVR: liquid crystal variable retarder, BS: beam splitter, PM: plane mirror, MS: mechanical stage, PS: piezoelectric stage, OB: objective, and DVA: differential voltage amplifier. The dimensions are not to scale.

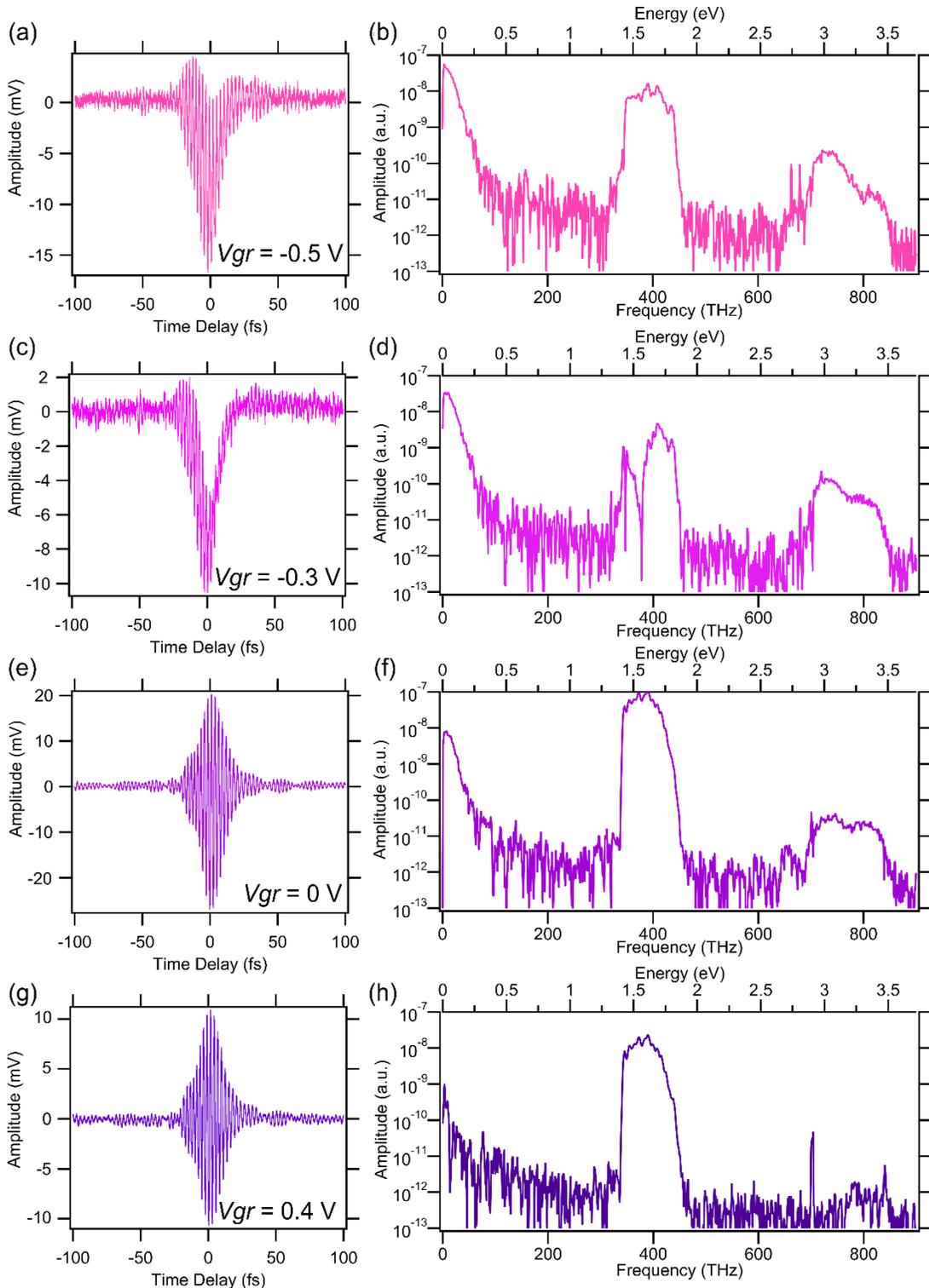

**Figure 3. $V_{gr}$ dependence of time domain signal. (a,c,e,g)** $V_{gr}$ dependence of the G/LAO/STO time domain signal. A DC offset-dependent splitting of the time domain signal appears at $V_{gr} = -0.3$ V. (T = 10 K, $V_{SD} = -1$ V.) **(b,d,f,h)** Power spectra of the time domain signals. For $V_{gr} = -0.3$ V, a sharp dip appears in the LNR response of the G/LAO/STO device at 380 THz. The curves are distinguished by color and all plots share the same color correspondences. See **Supplementary Figures S3-5** for more data from this experiment.

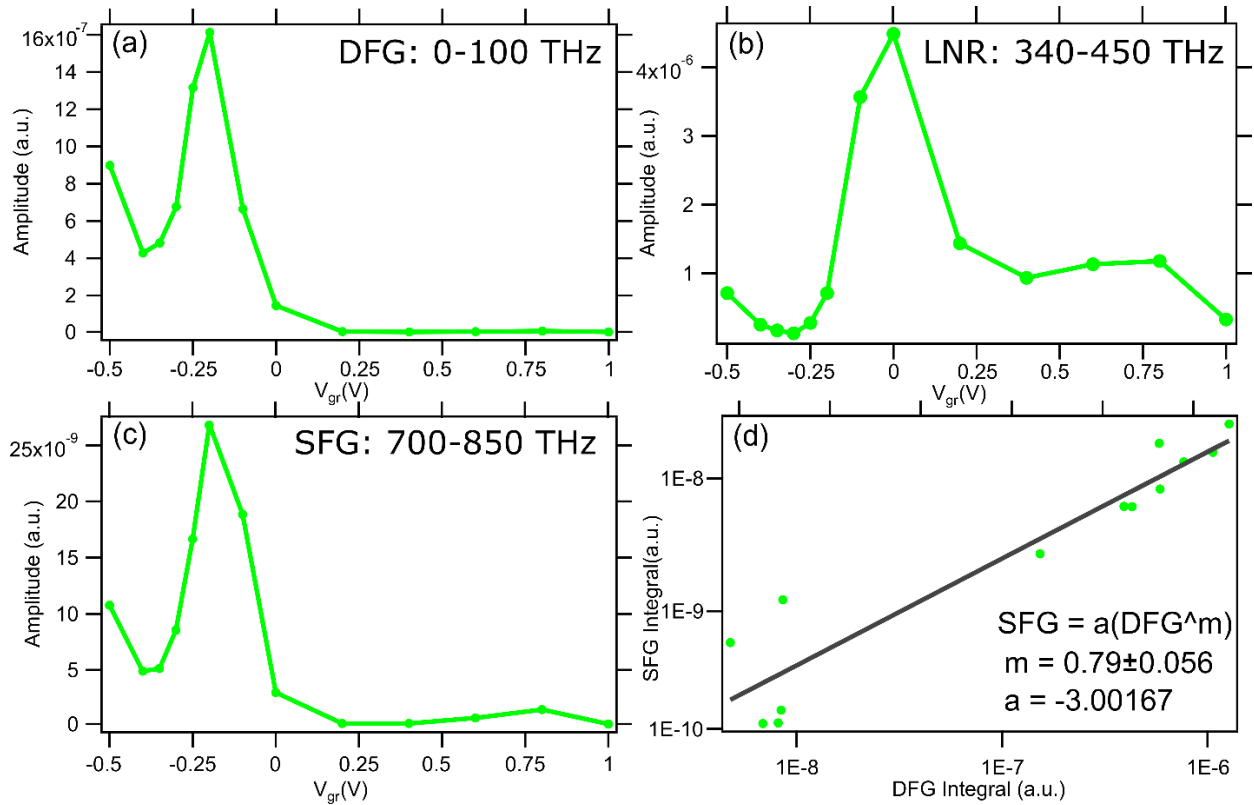

**Figure 4. Power spectra integrals.** Integrals of the **(a)** difference frequency generation (DFG) response of the device and **(b)** linear (LNR) and **(c)** second harmonic generation (SFG) response. Each data point represents a graphene DC offset value $V_{gr}$ at which the time domain measurement was taken. The DFG and SFG responses are maximal at $V_{gr} = -0.2$ V , close to where the sharp extinction feature appears, whereas the LNR response is maximal around a higher value of $V_{gr} = 0$ V. **(d)** Log-log plot of the DFG and SFG integral values with a power law fit to determine the power law relationship between them, which is equal to the slope, 0.79.

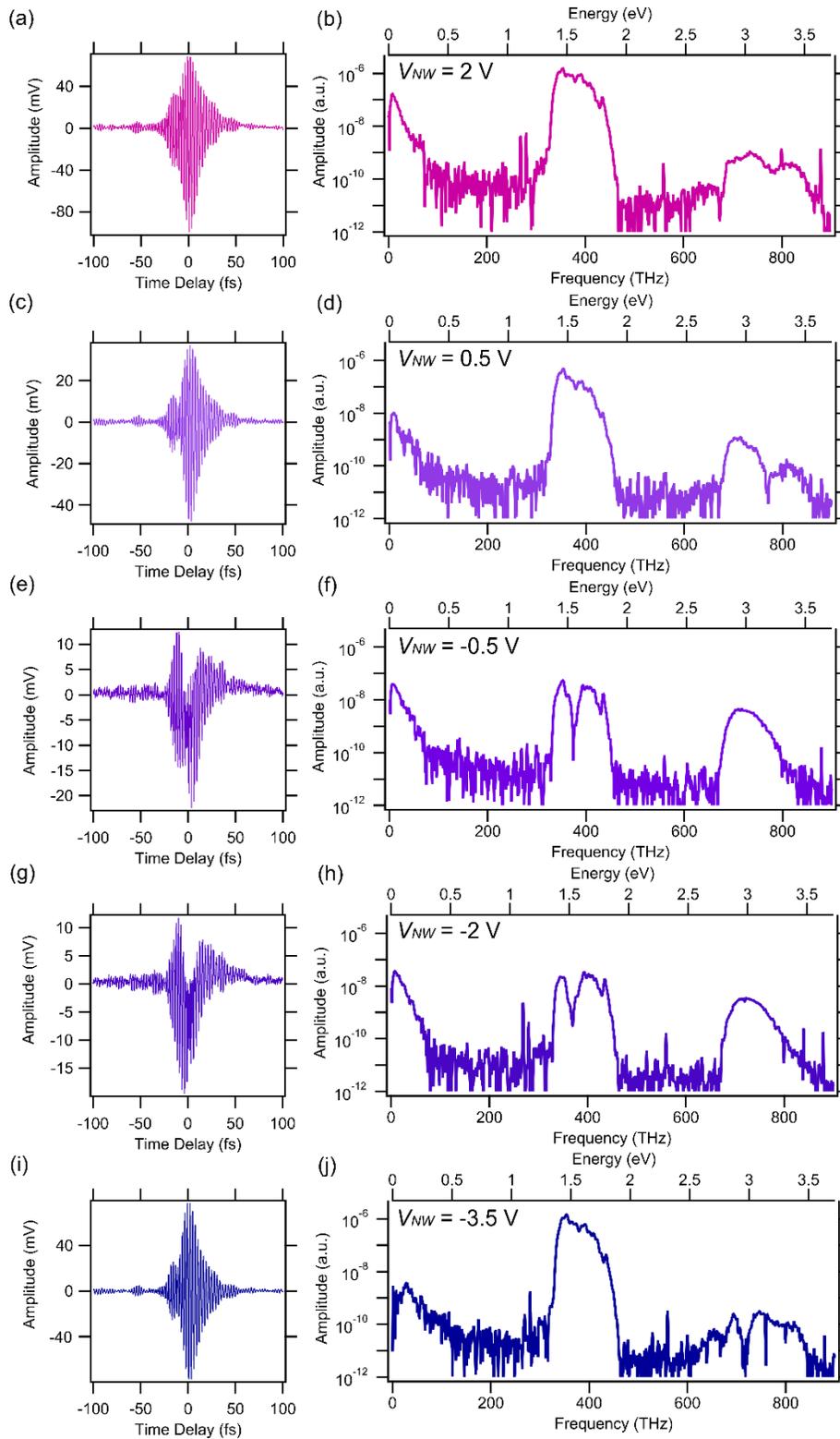

**Figure 5. $V_{NW}$ dependence of time domain signal.** Nanowire gating experiment. **(a,c,e,g,i)** Time domain signals at different $V_{NW}$ values. A gate-dependent splitting of the time domain signal appears between $V_{NW} = -0.5$ V, $-2.5$ V. (T = 45 K, $V_{SD} = -2$ V.) **(b,d,f,h,j)** Power spectra of the time domain signals. For $V_{NW} = -0.5$ V, a sharp extinction line appears in the NIR response at 375 THz. For $V_{NW} = 0.5$ V, $-3.5$ V, a sharp extinction line appears in the SFG response at 765 THz and 724 THz, respectively.

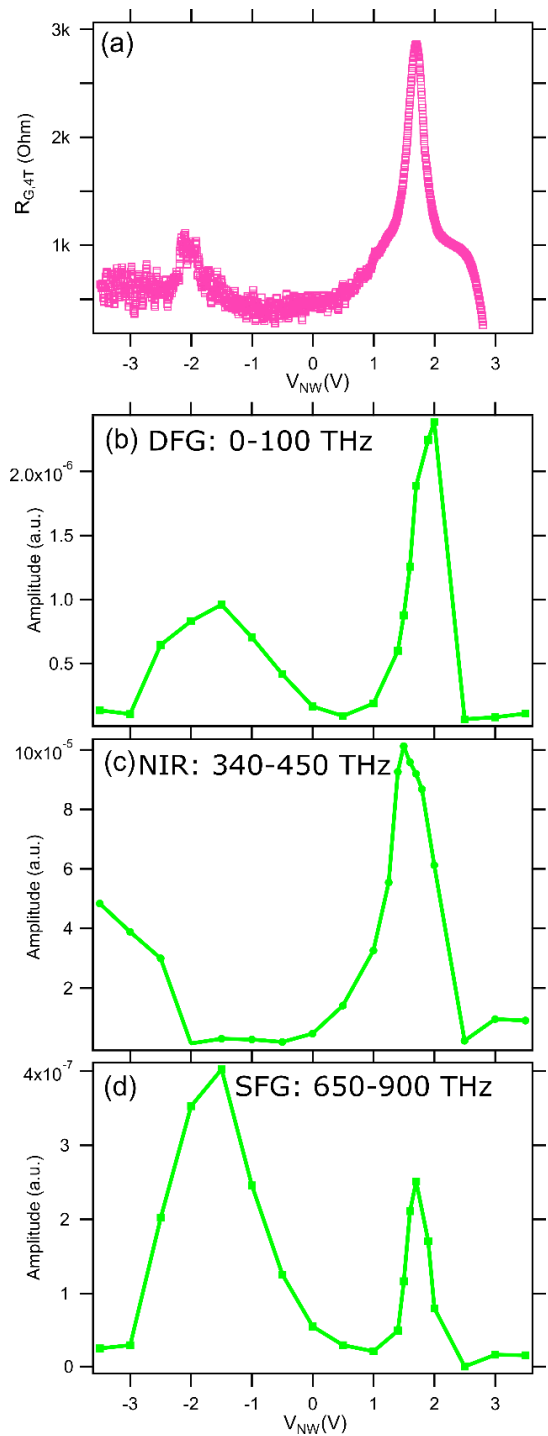

**Figure 6. Graphene resistance and power spectrum integrals. (a)** Four-terminal graphene resistance $R_{G,4T}$ measured as a function of $V_{NW}$ for the experiment shown in **Figure 5**. The presence of a second local maximum in $R_{G,4T}$ may be a signature of inhomogeneous doping in the graphene sheet. Integrals of the **(b)** DFG **(c)** LNR and **(d)** SFG responses as a function of $V_{NW}$. The nonlinear responses each have two maxima, one at $V_{NW} = -1$ V, where the LNR extinction line appears, and another at $V_{NW} = 1.7$ V, at the CNP. The linear response appears to only have a maximum at $V_{NW} = 1.7$ V.

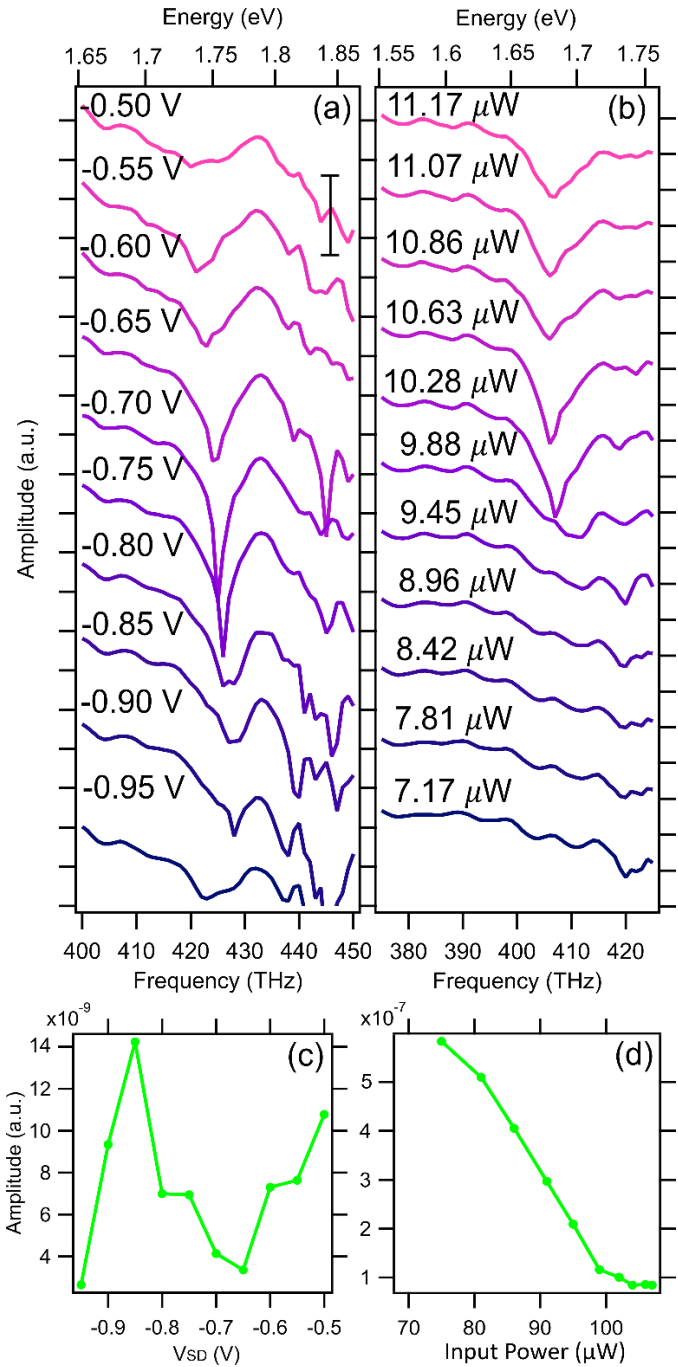

**Figure 7**. $V_{SD}$ **and power dependence of extinction feature.** Power spectra of time domain signals for **(a)** $V_{SD}$ dependence of the sharp extinction line ($T = 10$ K, $V_{gr} = 25$ mV) and **(b)** Input power dependence of the sharp extinction line ($T = 10$ K, $V_{SD} = -1.25$ V, $V_{gr} = 625$ mV.) Scale bar marks two orders of magnitude. **(c)** Integral of the power spectra in (a) from 420-430 THz as a function of $V_{SD}$. **(d)** Integral of the power spectra in (b) from 385-415 THz as a function of input power.

# Gate-Tunable Optical Nonlinearities and Extinction in Graphene/LaAlO$_3$/SrTiO$_3$ Nanostructures

*Erin Sheridan,*[1,2] *Lu Chen,*[1,2] *Jianan Li,*[1,2] *Qing Guo,*[1,2] *Hyungwoo Lee,*[3] *Jung-Woo Lee,*[3] *Chang-Beom Eom,*[3] *Patrick Irvin,*[1,2] *Jeremy Levy*[1,2*]

[1]Department of Physics and Astronomy, University of Pittsburgh, Pittsburgh, Pennsylvania 15260, USA

[2]Pittsburgh Quantum Institute, Pittsburgh, Pennsylvania 15260, USA

[3]Department of Materials Science and Engineering, University of Wisconsin-Madison, Madison, Wisconsin 53706, USA

[*]Correspondence: jlevy@pitt.edu

CONDUCTIVE AFM LITHOGRAPHY

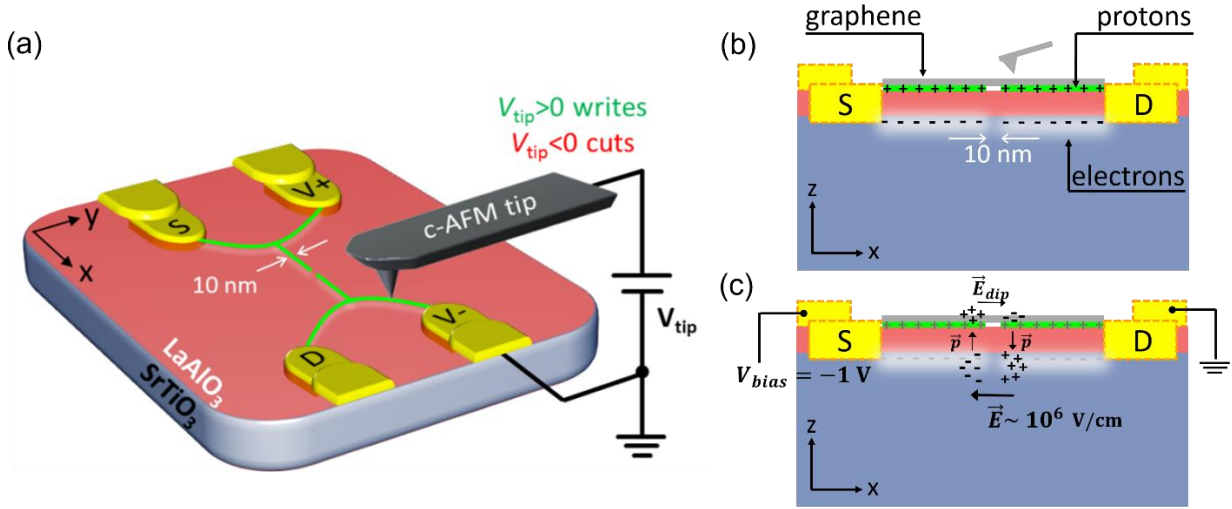

**Figure S1. (a)** Conductive atomic force microscope (c-AFM) lithography. Gold electrodes are patterned via conventional photolithography to form direct contact to the LAO/STO interface. The green wires represent the designed geometry for a typical nanojunction device. A positively biased AFM tip writes the conducting nanowires in contact mode, while a negatively biased AFM tip creates a nanojunction by cutting across the nanowire. **(b)** A side view of the sample shows that the c-AFM-lithography-defined device is located beneath the graphene at the interface of the LAO/STO heterostructure. Both the nanowires and the nanojunction have a spatial confinement of approximately 10 nm. The dimensions are not to scale. **(c)** Side view depicting the strong electric field across the LAO/STO nanojunction for $V_{SD} = -1$ V and the corresponding dipole electric field $\vec{E}_{dip}$ that is produced in the graphene to screen it.

GRAPHENE FOUR-TERMINAL RESISTANCE MEASUREMENT

To measure the graphene four terminal resistance $R_{G,4T}$, we source 1 mV of AC voltage at 2.37 Hz at the source electrode (labeled I+ in **Figure S2**) and ground the drain electrode (I-). Voltage sensing leads (V+/-) are used for a four-terminal voltage measurement. $R_{G,4T}$ is calculated as $R = \frac{(V_+ - V_-)}{I_-}$ and equivalently, the conductance is calculated as $G = \frac{I_-}{(V_+ - V_-)}$.

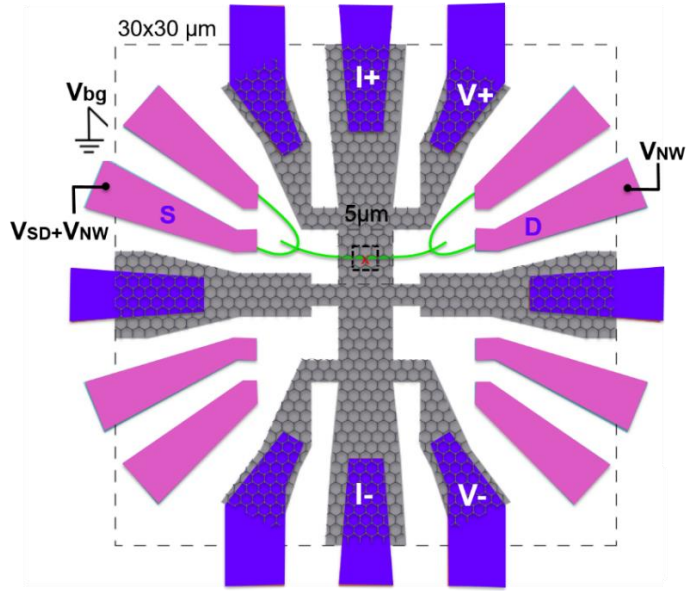

**Figure S2. Graphene resistance measurement.** Top-down diagram of graphene/LAO/STO sample structure with graphene four terminal resistance measurement connections labeled.

ADDITIONAL DATA FROM $V_{gr}$ DEPENDENCE EXPERIMENT

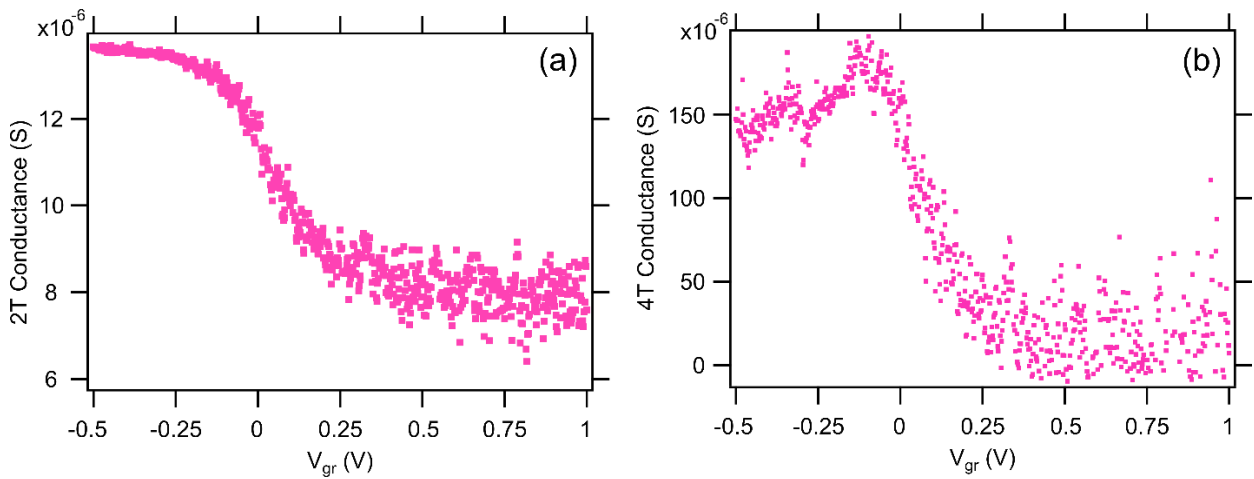

**Figure S3. Graphene conductance vs. $V_{gr}$.** (a) Two-terminal and (b) four-terminal conductance curves for graphene as a function of the global graphene gate $V_{gr}$ for the experiment described in main text

**Figures 1, 3, 4**. As compared to the nanowire gate $V_{NW}$ (see main text **Figure 6**), a $V_{gr}$ sweep is not able to resolve the Dirac point as clearly.

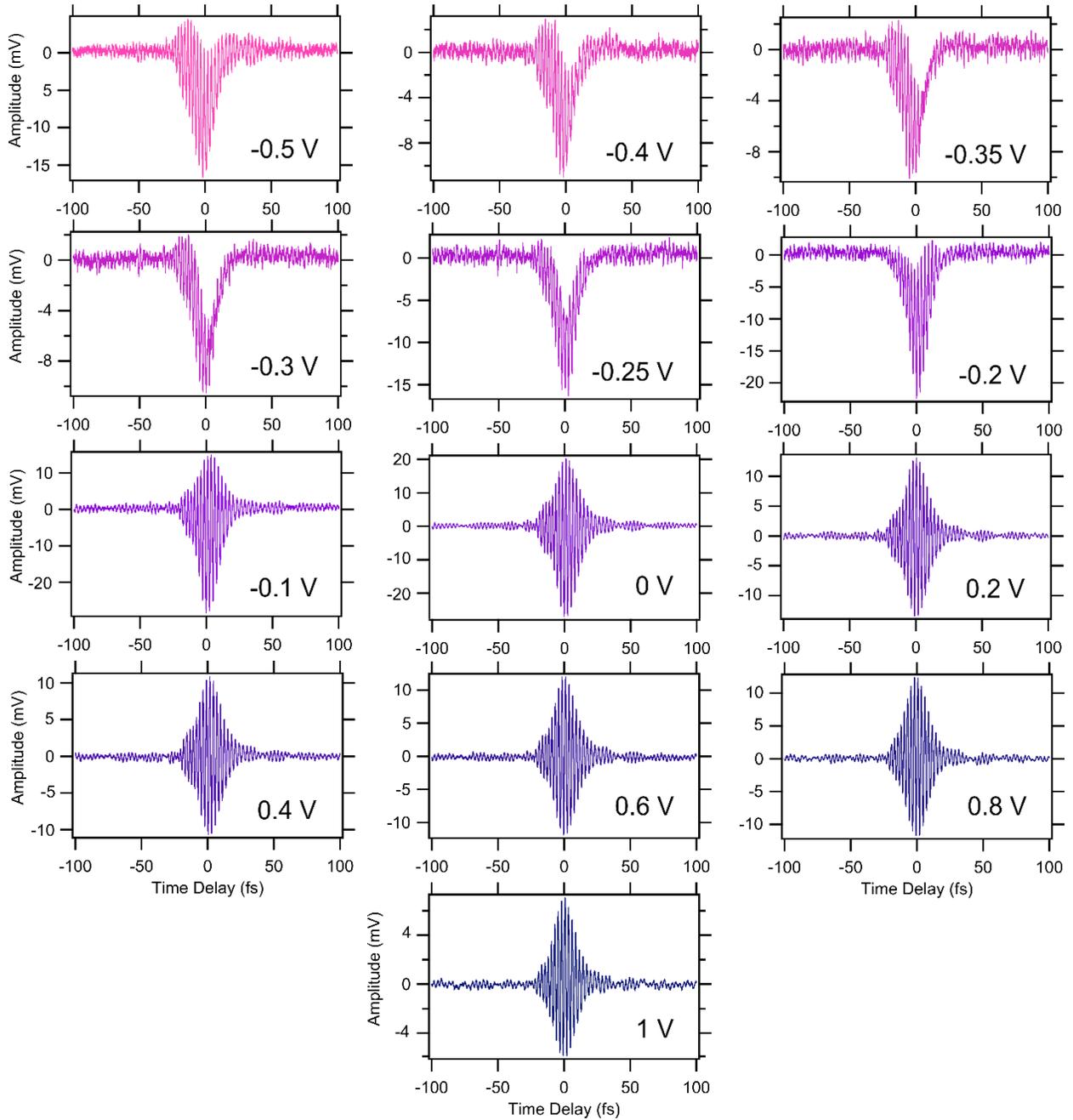

**Figure S4**. Time domain measurements from $V_{gr}$-dependent experiment discussed in the main text. Note that the splitting feature in the time domain signal is only observed between $V_{gr}$ values of $-0.5$ V and $-0.2$ V. Note also that the nonlinearity of the time domain signal diminishes at $V_{gr}$ values above $-0.1$ V.

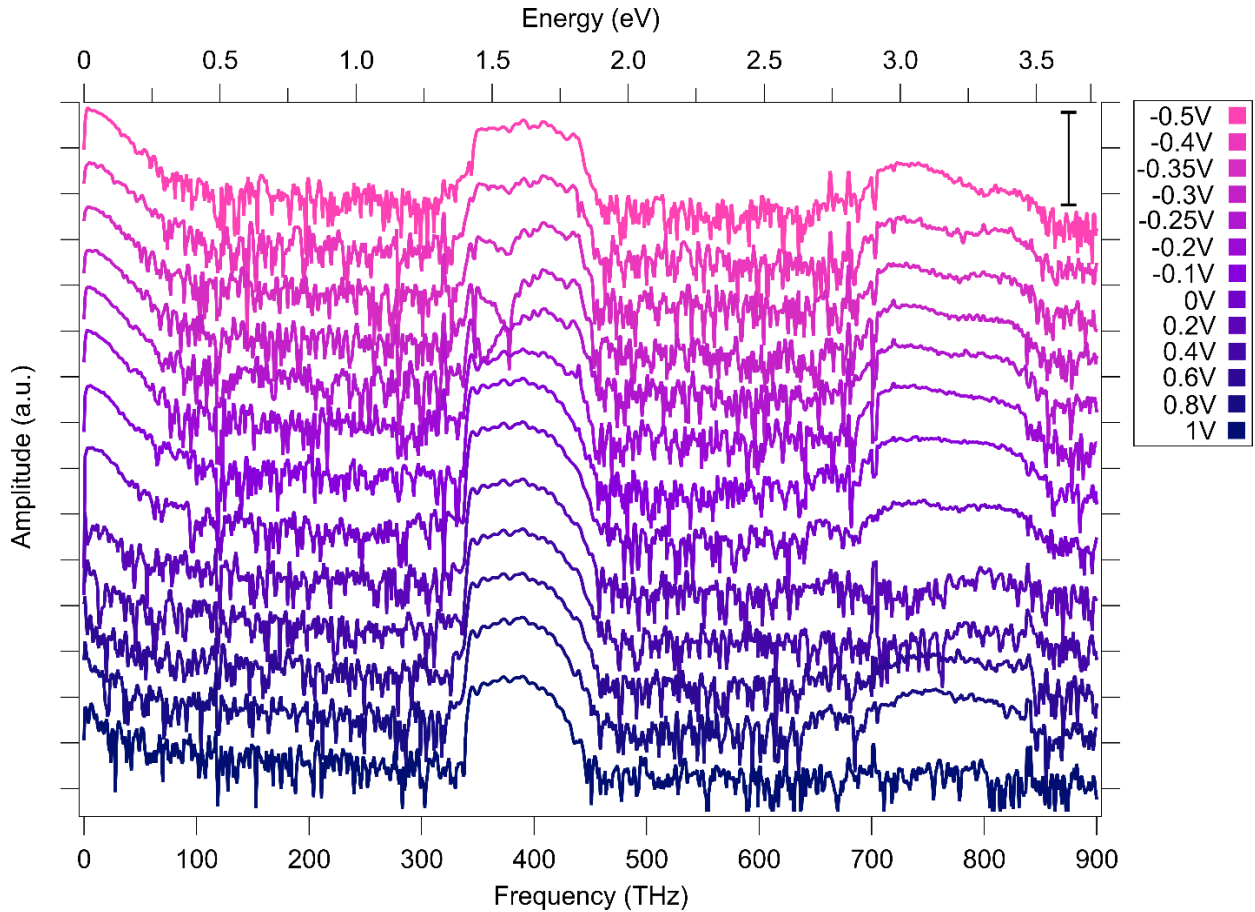

**Figure S5**. Power spectra from $V_{gr}$ experiment in the main text. Plots are vertically offset for clarity. Scale bar marks four orders of magnitude.

ADDITIONAL EXPERIMENTS

LAO/STO NANOJUNCTION SOURCE-DRAIN BIAS AND OPTICAL POWER DEPENDENCE

The interplay between the $V_{SD}$ and optical power dependence of the extinction feature is investigated by constructing a "matrix" of data by repeating the same optical power sweep at different $V_{SD}$ values. The input optical power to the G/LAO/STO nanostructure is decreased from 9.42 µW to 7.88 µW (a 16% change) at $V_{SD} = -0.2$ V, then the magnitude of $V_{SD}$ is increased in 50 mV steps to $-0.4$ V. As the $V_{SD}$ value increases, the power-dependent power spectra shift downwards, or lower in power. Black arrows in **Figure S6** mark the first power spectrum for which two extinction features are visible in the LNR response. The downward shift in the black arrows shows how the power dependence shifts with increasing $V_{SD}$.

Though the interplay between optical power and bias is complex and we cannot obtain an exact scaling law depicting their relationships, the two parameters appear to act together, as powerful knobs to fine-tune extinction feature frequencies and extinction ratios.

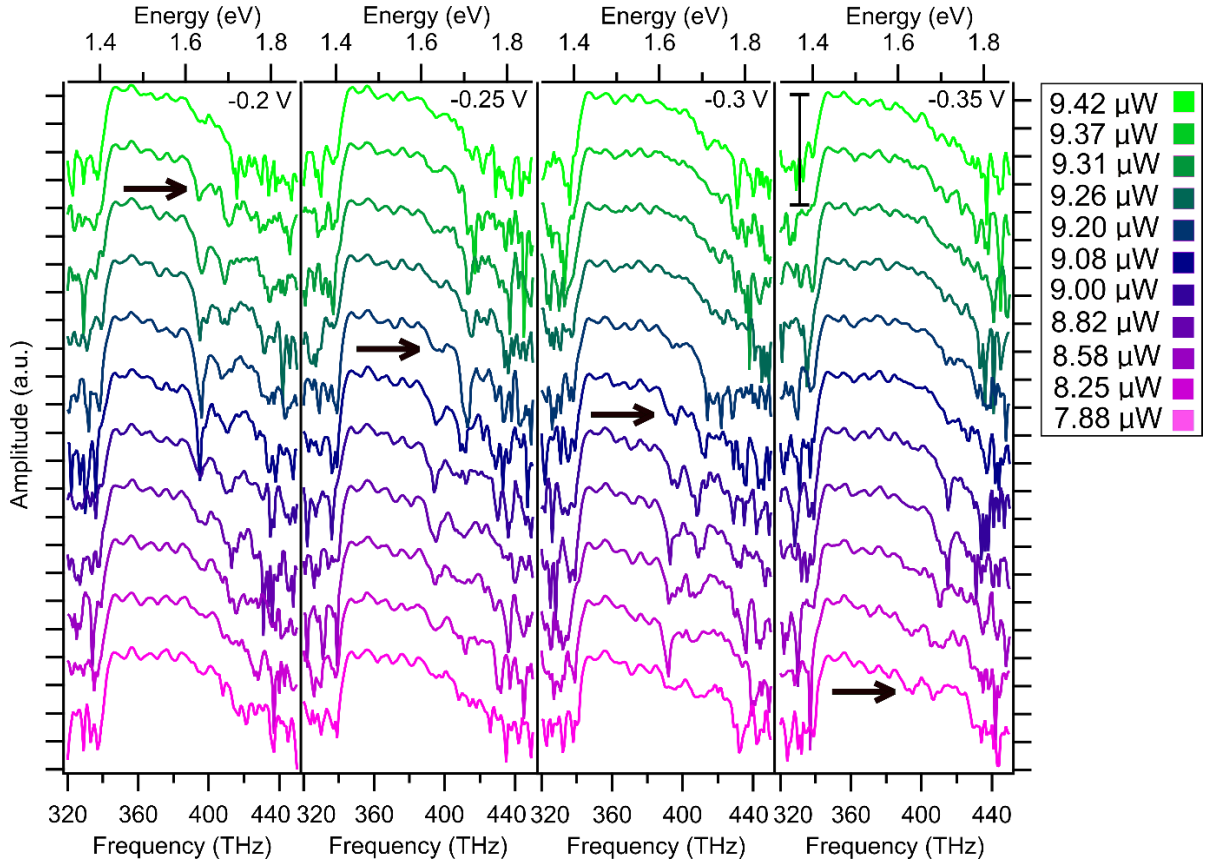

**Figure S6**. Power spectra from four power dependence experiments at different $V_{SD}$ values. Each column of data represents a power-dependent experiment at the labeled $V_{SD}$ value. The scale bar marks four orders of magnitude. Black arrows mark the first power spectrum where two extinction features are visible.

LINEAR POLARIZATION DEPENDENCE OF EXTINCTION FEATURE

To study the polarization dependence of the VIS-NIR sharp absorption line, we place a thermally-stabilized liquid crystal variable retarder optic (LCVR) in the beam path directly before the Michelson interferometer. The LCVR can act as a broadband half waveplate. This allows us to switch the linear polarization of the input pulses from 0 degrees (parallel to the device) to 90 degrees (perpendicular to the device) by changing the voltage applied to the liquid crystal. We can therefore instantaneously switch between two linear polarizations without disturbing the beam path, as we would if we manually rotated a half wave plate.

For each experiment, a measurement is taken at 0 degrees polarization, or parallel to the nanojunction device. The LCVR is then switched to 90 degrees polarization and another measurement is taken. In the two experiments shown in **Figure S7**, the linear polarization is switched from 0 degrees to 90 degrees, then immediately back to 0 degrees, then 90 degrees again. For both of the experiments shown, the nanostructure appears to have two different resonances; one when the light is polarized along the nanowire, and another when the light is polarized perpendicular to it.

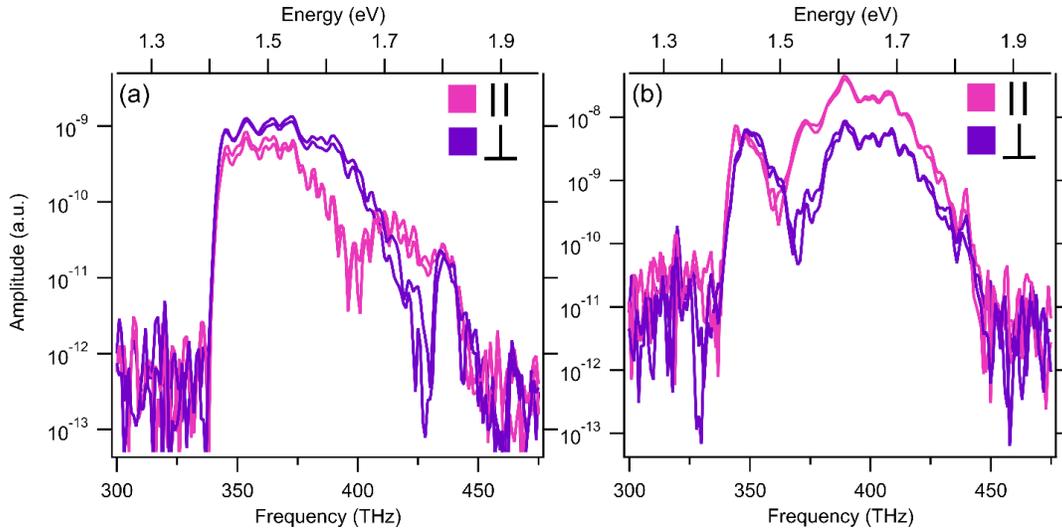

**Figure S7.** Polarization dependence of sharp extinction feature. Pink: light polarized parallel to nanojunction device (parallel to nanowire containing the nanojunction.) Purple: light polarized perpendicular to device. Power spectrum at each polarization for **(a)** T = 5 K, $V_{SD} = -1$ V, $V_{gr} = 1$ V and **(b)** T = 10 K, $V_{SD} = -2.3$ V, $V_{gr} = 2.5$ V. At perpendicular polarizations, extinction features appear at different frequencies. For (a), the extinction appears at about 400 THz for parallel polarization and at 428 THz for perpendicular polarization. For (b), the extinction appears at 359 THz for parallel polarization and at 368 THz for perpendicular polarization.

ADDITIONAL EXAMPLES OF EXTINCTION FEATURES

Three examples of sharp extinction features are shown in **Figure S8**. These features appear at different frequencies across the LNR range defined by the input excitation, and at different LAO/STO nanojunction bias $V_{SD}$ or $V_{gr}$ values, depending on the individual device. Extinction features were observed on 14 devices across 5 different samples and sheets of graphene.

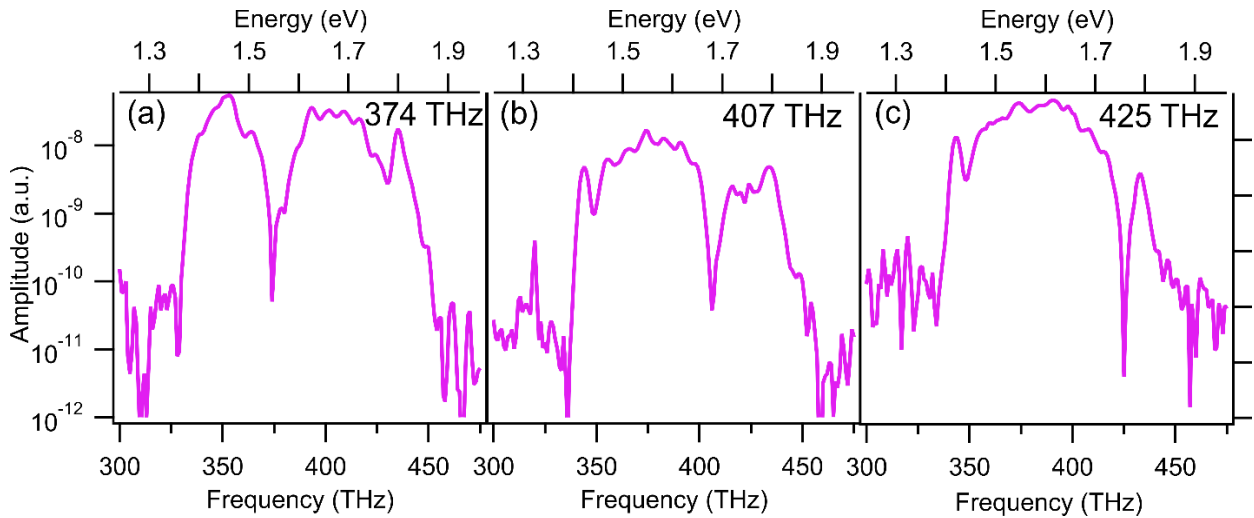

**Figure S8.** Extinction features in three different experiments. **(a)** T = 45 K, $V_{SD} = -2$ V, $V_{NW} = -0.3$ V **(b)** T = 10 K, $V_{SD} = -1.25$ V, $V_{gr} = 0.625$ V. **(c)** T = 10K, $V_{SD} = -0.7$ V, $V_{gr} = -25$ mV.

CONTROL EXPERIMENTS with LAO/STO NANOJUNCTIONS

To verify that the observed sharp extinction features originate from the graphene in the G/LAO/STO nanostructure and not from the LAO/STO, a nanojunction device was written on a LAO/STO sample without graphene. A nanowire gating experiment is performed by changing $V_{NW}$. Time domain measurements (**Figure S9**) are taken at a constant $V_{SD} = -1$ V while changing $V_{NW}$ and the corresponding power spectra are shown in **Figure S10**. Without graphene, the device shows no observable nanowire gate dependence. No sharp extinction feature is seen in the NIR or SFG regions and there is no noticeable change in the nonlinear THZ or SFG response with $V_{NW}$.

While we are unable to know exactly which components of the THZ and SFG responses come from the graphene and which come from the LAO/STO nanojunction, LAO/STO without graphene clearly has a characteristic nonlinear response that is not gate-dependent. On the other hand, nanowire gating of the G/LAO/STO nanostructure reveals dramatic changes in both the THZ and SFG responses. We therefore assume that the observed nonlinearities in the G/LAO/STO nanostructures are comprised of a combination of both the graphene and LAO/STO nonlinear responses, and the gate dependence originates in the graphene response.

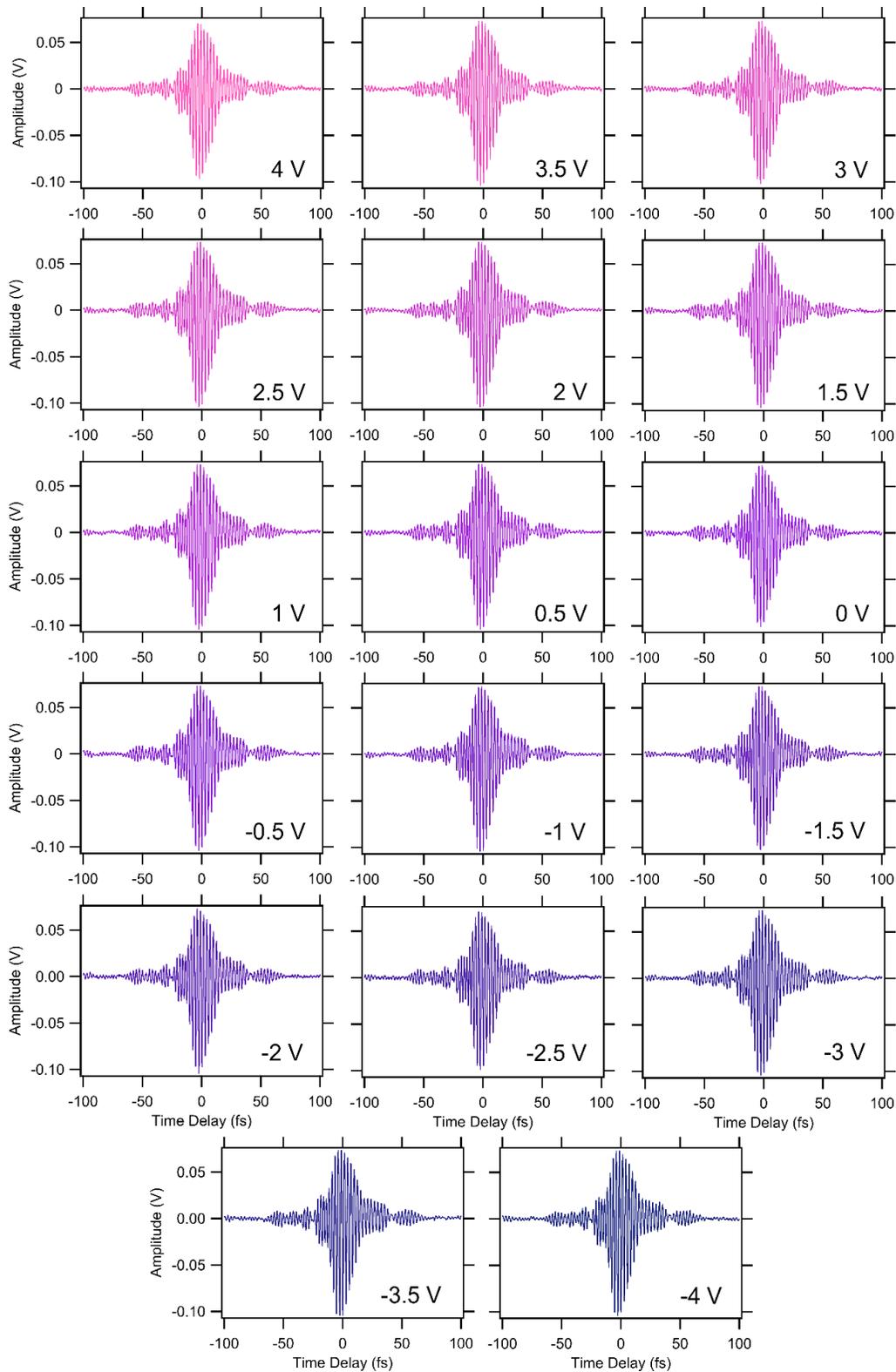

**Figure S11** shows the integrals of the relevant regions of the power spectra in **Figure S10**. Note that, unlike the G/LAO/STO nanostructure, the linear and nonlinear responses of LAO/STO nanojunction without graphene do not show any dependence on the nanowire gate value.

**Figure S9**. Time domain signals from control interface gating experiment. The signal at each $V_{int}$ value is nearly identical to the rest; no splitting features appear, and there are no noticeable changed in amplitude.

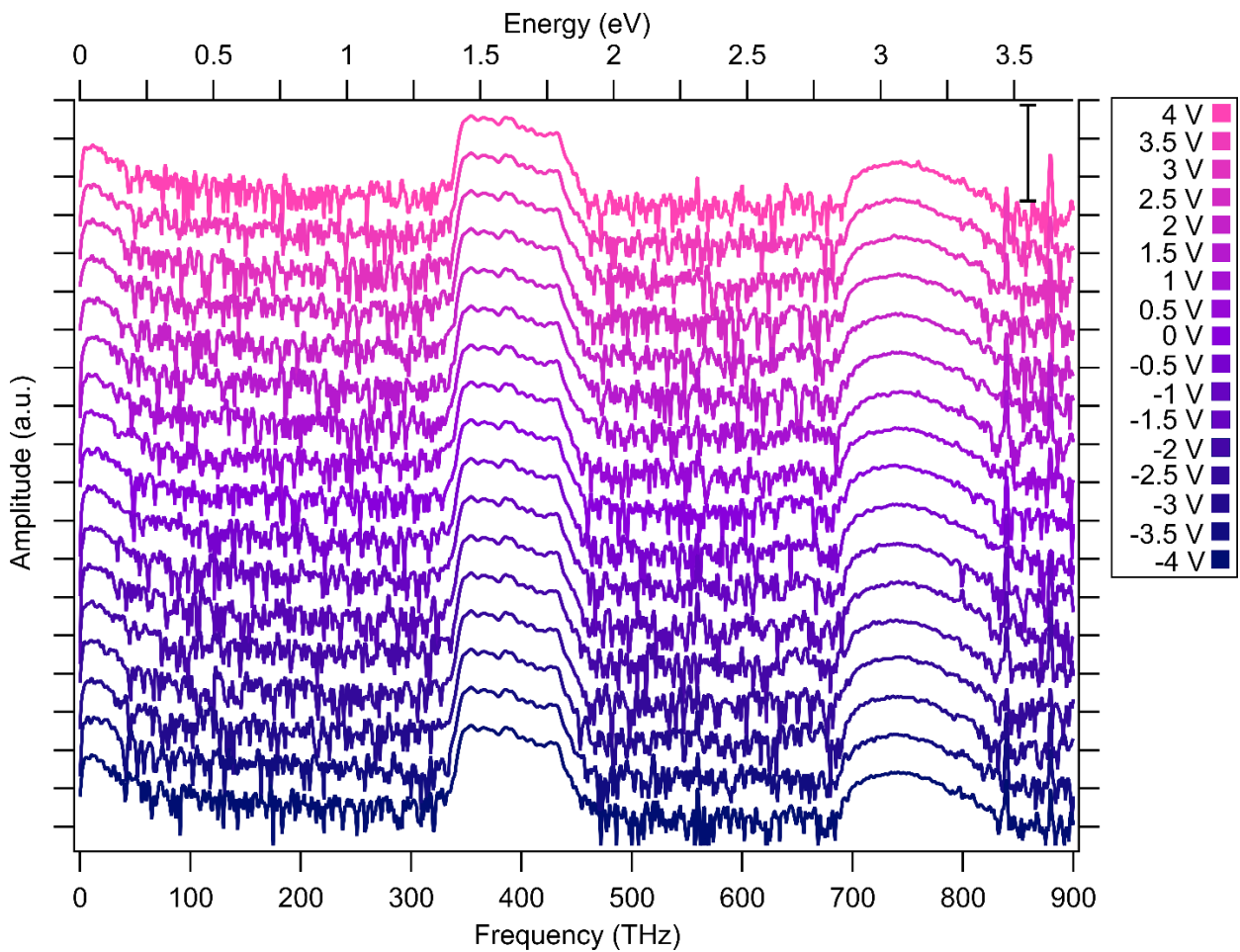

**Figure S10**. Power spectra of time domain signals in **Figure S9**. Different power spectra are offset vertically for clarity. Scale bar denotes five orders of magnitude. No $V_{NW}$ dependence or extinction feature in the NIR or SHG region is observed.

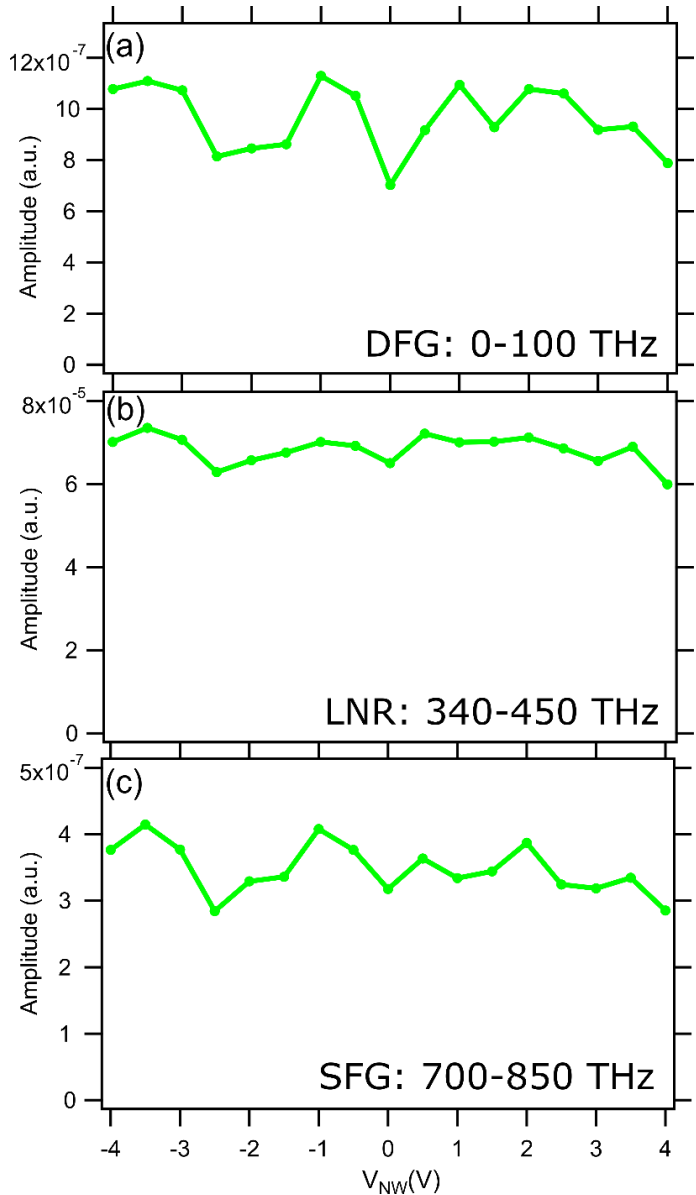

**Figure S11**. Integrals of the power spectra in **Figure S10** as a function of the nanowire gate $V_{NW}$ for the **(a)** THZ **(b)** NIR and **(c)** SHG region. Notice that the linear and nonlinear responses of the LAO/STO nanojunction without graphene do not show any observable gate dependence.

GRAPHENE FOUR TERMINAL RESISTANCE CURVES vs. $V_{SD}$

To learn more about the inhomogeneous gating of the graphene by $V_{NW}$ and $V_{SD}$, we measured $R_{G,4T}$ as a function of $V_{NW}$ at different $V_{SD}$ values. If the biased LAO/STO nanojunction is a source of inhomogeneous doping in the graphene sheet, we should expect to see a separation of two Dirac point features as the LAO/STO nanojunction source-drain bias $V_{SD}$ increases in magnitude. As $V_{NW}$ is tuned, different regions of the graphene sheet will be gated to the CNP at different gate values, and the difference in these two gate values should increase as the inhomogeneity of the gating increases.

As shown in **Figures S12-13**, preliminary experiments match our prediction. A plot of the Dirac point separation vs. $V_{SD}$ shows a clear increase in peak separation as the magnitude of $V_{SD}$ increases. However, the CNP appears quite broad, which obscures the clarity of the results to some extent. Follow-up experiments are required to further clarify the nature of inhomogeneous doping of the graphene by the nanojunction.

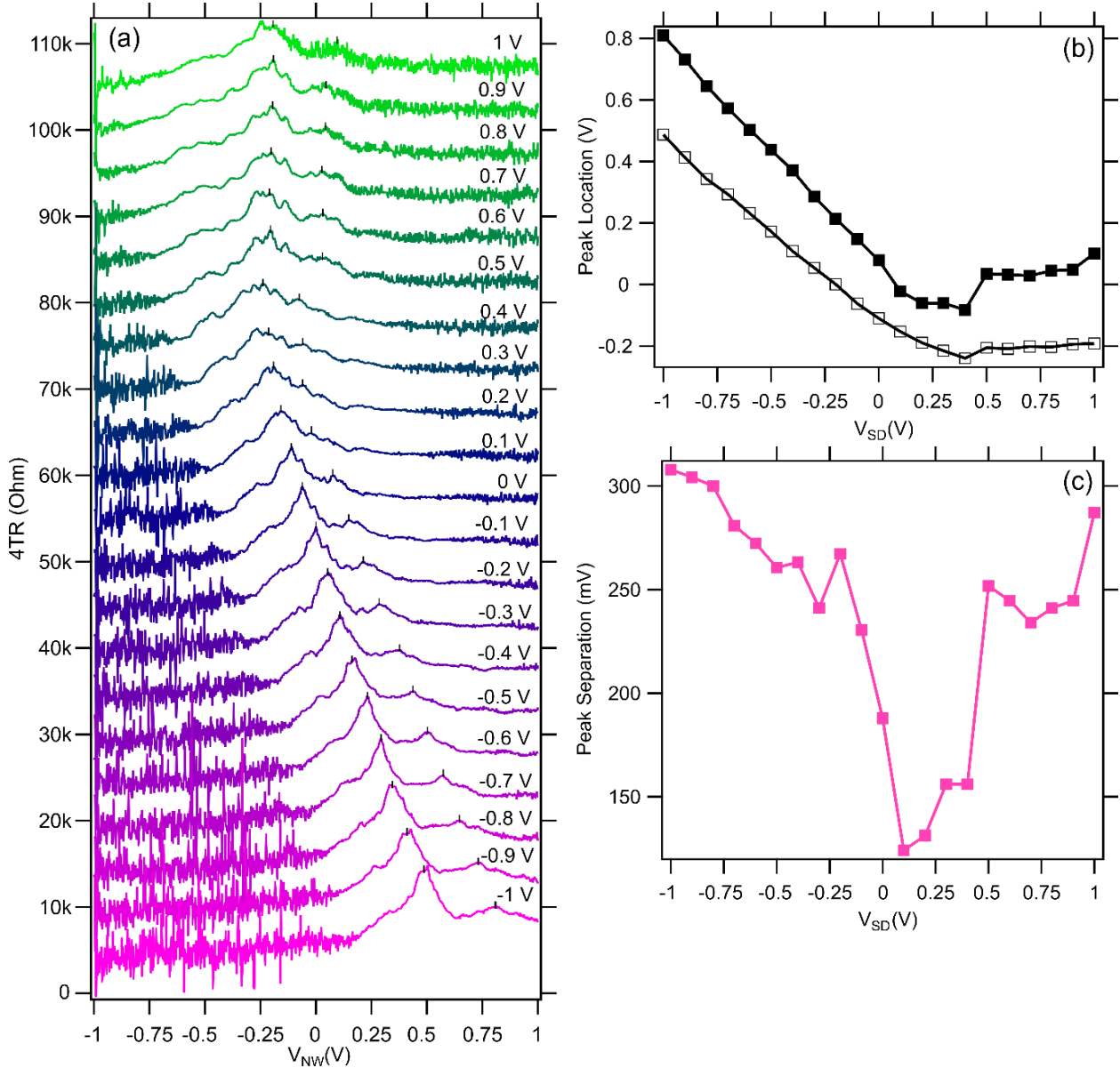

**Figure S12**. **Dirac feature separation. (a)** $R_{G,4T}$ as a function of $V_{NW}$ for different differential gate $V_{SD}$ values at T = 20 K and 9 µW input power. Resistance curves are offset by 5 kOhm for clarity and labeled with their corresponding $V_{SD}$ value. Small black ticks mark the gate value location of the maximum. **(b)** Location of the two observed resistance peaks for each $V_{SD}$ value. **(c)** Peak separation of two Dirac point maxima in $R_{G,4T}$ curves vs. $V_{SD}$. Notice that around $V_{SD} = 0.1$ V, the peak separation is minimal, while the separation grows as the magnitude of $V_{SD}$ increases.